# The Accuracy of Restricted Boltzmann Machine Models of Ising Systems

David Yevick and Roger Melko
Department of Physics
University of Waterloo
Waterloo, ON N2L 3G7

**Abstract:** Restricted Boltzmann machine (RBM) provide a general framework for modeling physical systems, but their behavior is dependent on hyperparameters such as the learning rate, the number of hidden nodes and the form of the threshold function. This article accordingly examines in detail the influence of these parameters on Ising spin system calculations. A tradeoff is identified between the accuracy of statistical quantities such as the specific heat and that of the joint distribution of energy and magnetization. The optimal structure of the RBM therefore depends intrinsically on the physical problem to which it is applied.

**Keywords:** Statistical methods, Ising model, Machine learning algorithms

**Introduction:** Numerical methods for analyzing lattice models can be classified as either stochastic or deterministic. Stochastic procedures employ random number generation to sample appropriate regions of the configuration space from which macroscopic quantities of interest are evaluated. The specification of these regions has become increasingly sophisticated, progressing from unbiased Monte-Carlo procedures to Markov-based procedures such as the Metropolis, [1] multicanonical, [2] [3] [4] and Wang-Landau [5] [6] [7] techniques. The latter three methods preferentially visit low probability regions of thermodynamic (e.g. macroscopic) system variables, enabling efficiency increases of many orders of magnitude in the calculation of statistically rare events. Such algorithms can be further refined by aggregating the statistics of both the accepted and rejected Markov chain transitions into a matrix whose lowest order eigenvector corresponds to the unbiased probability distribution function (density of states) [8] [9] [10] [11] [12] [13]. Unfortunately, however, particularly in the presence of phase transitions, correlations between successive samples limit the accuracy of stochastic approaches. Although refinements such as the monitored sampling of appropriate configuration space regions [12] and cluster reversal algorithms [14] can significantly decrease computation times, a large number of samples are still required to achieve acceptable results for quantities such as the specific heat that involve multiple derivatives of the partition function.

In contrast, deterministic methods are typically unaffected by sample correlations. These include analytic procedures for small systems as well as approximate renormalization group and tensor network algorithms [15] [16] [17] [18] [19] [20] [21] [22] [23] [24]. These latter procedures iteratively transform in an approximate yet controllable manner quantities defined on extended or infinite lattices into analogous quantities on increasingly smaller lattices. However, unlike stochastic methods, which trivially adapt to grids of any geometrical structure, deterministic methods such as the renormalization group cannot be simply applied to heterogeneous networks since decimating the network will then not generate a network of the same structure.

Recently, machine learning and stochastic algorithms have been combined by several authors. In [25] [26] a lattice model is mapped onto a smaller neural network model that, similarly to cluster algorithms, rapidly generates global steps near the critical temperature. Alternatively, in [27] a method inspired by the RBM was proposed that not only interpolates among various existing cluster reversal procedures but also suggests novel analogous techniques. Here, however, a straightforward implementation of machine learning is considered in which a RBM is trained with realizations from a Metropolis sampling calculation of a two-dimensional Ising spin system at a fixed temperature. [28] [29] [30] Subsequently, the machine generates new samples from which statistical properties such as the specific heat are determined. Such an approach can generate models that, despite possessing fewer nodes than the number of lattice spins, still reproduce the salient features of phase transitions. In contrast to previous studies, however, the accuracy of the RBM procedure will be quantified by considering not only the specific heat but also the distribution of states in energy-magnetization space at a temperature for which the system exhibits large fluctuations in magnetization. This analysis, which previously proved instrumental in the design of novel monitoring and control procedures for biased sampling, [12] [14] yields novel insight into the behavior of the RBM.

**Computational Methods:** While the notation and equations that describe the RBM implementation of the Ising model are not immediately evident, the underlying concepts can be simply clarified in the context of a streamlined but efficient MATLAB© program with the syntax conventions of [31] [32]. The analogous code in TensorFlow was found to be both more complex and somewhat less efficient.

The first relevant code line, namely **load realizationSaveFile** reads data from a file containing, in the present case, **trainingDataRCLength =** 600,000 successive realizations of a $8 \times 8$ Ising spin system generated by a standard Markov chain Metropolis sampling program at a temperature $T$ and outputs a 3-dimensional tensor of size $8 \times 8 \times 600,000$. The tensor is subsequently flattened to a two-dimensional matrix of size $600,000 \times 64$ (note the transpose operator **'** below) and the $-1$ spin down values are replaced with 0 by

**trainingDataRC = reshape( realizationSave, 64, size( realizationSave, 3 ) )';**
**trainingDataRC = max( trainingDataRC, 0 );**

The restricted Boltzmann machine (RBM) is here comprised of an array of **numberOfVisibleNodes** "visible" nodes, each of which corresponds to one of the input spin variables. The *m*:th visible node is coupled to every member, *n*, of **numberOfHiddenNodes** "hidden" nodes with a weight given by **weightMatrixRC(m, n).** Assigned to the *m*:th visible node and to the *n*:th hidden node are the "biases" **visibleBiasesR(m)**, and **hiddenBiasesR(n)**, respectively. Optimally, the learning phase of the program employs the flattened data set to generate weights and biases such that the trained RBM transforms appropriate inputs into new visible values that both distribute over configuration space similarly to the original training set and further accurately reproduce macroscopic quantities of interest such as the specific heat. Initially, random values are initially assigned to the biases and weights:

**weightMatrixRC = ( upperBiasValue - lowerBiasValue ) * …**
    **rand( numberOfVisibleNodes, numberOfHiddenNodes ) + lowerBiasValue;**
**visibleBiasesR = ( upperBiasValue - lowerBiasValue ) * …**
    **rand( 1, numberOfVisibleNodes ) + lowerBiasValue;**
**hiddenBiasesR = ( upperBiasValue - lowerBiasValue ) * …**
    **rand( 1, numberOfHiddenNodes ) + lowerBiasValue;**

To minimize the correlation between successive Metropolis realizations, the training set elements are input into the RBM in a random order through a permuted index vector.

**myIndices = randperm( trainingDataRCLength );**

Two procedures for the RBM are commonly employed for the training phase, both of which will be considered in this paper. The first of these is implemented as follows:

**hiddenValuesR = zeros( 1, numberOfHiddenNodes );**
**visibleValuesProposalR = zeros( 1, numberOfVisibleNodes );**
**hiddenValuesProposalR = zeros( 1, numberOfHiddenNodes );**

**for loop = 1 : numberOfRealizations**

   **visibleValuesR = trainingDataRC(myIndices(loop), :);**
   **hiddenValuesR = generateHiddenProposal( visibleValuesR, weightMatrixRC, …**
      **hiddenBiasesR, numberOfHiddenNodes, inverseTemperature );**

   **positiveWeightGradientRC = visibleValuesR' * hiddenValuesR;**

   **visibleValuesProposalR = generateVisibleProposal( hiddenValuesR, weightMatrixRC, …**
      **visibleBiasesR, numberOfVisibleNodes, inverseTemperature );**
   **hiddenValuesProposalR = generateHiddenProposal( visibleValuesProposalR, …**
      **weightMatrixRC, hiddenBiasesR, numberOfHiddenNodes, inverseTemperature );**

   **negativeWeightGradientRC = visibleValuesProposalR' * hiddenValuesProposalR;**

   **weightMatrixRC = weightMatrixRC + learningRate * …**
      **( positiveWeightGradientRC – negativeWeightGradientRC );**
   **visibleBiasesR = visibleBiasesR + learningRate * ( visibleValuesR – visibleValuesProposalR );**
   **hiddenBiasesR = hiddenBiasesR + learningRate * …**
      **( hiddenValuesR – hiddenValuesProposalR );**

**end**

In particular, the elements of each randomly chosen realization are employed to set each visible node to a "neural excitation" value 0 or 1. These visible node values, $\vec{v}$, are passed to a function **generateHiddenPropopsal( )** together with the weights, $\mathbf{M}$, and hidden bias values, $\vec{b}_h$. This function,

**function hiddenProposalR = generateHiddenProposal( visibleValuesR, weightMatrixRC, …**
      **hiddenBiasesR, numberOfHiddenNodes, inverseTemperature )**
**computedHiddenProbabilitiesR = fermiFunction( visibleValuesR * weightMatrixRC + …**
      **hiddenBiasesR, inverseTemperature );**
**hiddenProposalR = zeros( 1, numberOfHiddenNodes );**
**hiddenProposalR(computedHiddenProbabilitiesR > rand( 1, numberOfHiddenNodes )) = 1;**

first generates a real value at each hidden node given by $\vec{h}_r = \vec{v}\mathbf{M} + \vec{b}_h$ and then calls an activation function, here a Fermi function modified to prevent numerical overflow or underflow. In this study, we further introduce an additional **inverseTemperature** parameter in order to adjust the rapidity of the 0 to 1 transition.

**function result = fermiFunction( x, inverseTemperature )**
**exponentCutoff = 20;**

```
exponentVariable = x * inverseTemperature;
myExponent = min( max( -exponentCutoff, exponentVariable ), exponentCutoff );
result = 1.0 ./ ( 1.0 + exp( -myExponent ) );
```

The Fermi function maps the vector $\vec{h}_r$ to a new vector **computedHiddenProbabilitiesR** with values ranging between 0 and 1. To transform this vector into a set of neural excitations, $\vec{h}$, for $i = 1, 2, \ldots.,$ **numberOfHiddenNodes,** a uniformly generated random number, $r_i$, is selected in the interval $[0,1]$, and the $i$:th element of the vector, $\left(\vec{h}\right)_i$, namely **hiddenProposalR(i)** is set to unity if $(h_r)_i > r_i$. Note that this prescription is stochastic and therefore generates different hidden vectors $\vec{h}$ for different random number sequences. Consequently, a meaningful solution only emerges after many input samples are processed.

Once a hidden vector is constructed, the procedure evaluates the outer (Helmholz) product **positiveWeightGradientRC** according to $\vec{v} \otimes \vec{h} \equiv \vec{v}^T \vec{h}$. Since prior knowledge of the weights and biases does not exist, the current values of these quantities can be considered the optimal basis for the recovery of the input spin distribution from the hidden variables. The $(i,j)$:th element of **positiveWeightGradientRC** thus is the best currently available indicator of the ideal coupling from input node $i$ to output node $j$. Hence it can be employed to slightly enhance the matrix element $(\mathbf{M})_{ij}$. However, the computed hidden nodes $\vec{h}$ together with the network variables normally differ from the optimal set. To compensate accordingly, a statistical realization of the visible node values predicted by $\vec{h}$ and $\mathbf{M}$ is obtained by computing $\vec{v}_r = \vec{h}\mathbf{M}^T + \vec{b}_v$ and then again comparing each of the values of $\vec{v}_r$ with a randomly generated uniformly distributed number on the unit interval. In the program above, this is performed by passing **hiddenProposalR** into a second function **generateVisibleProposal( )** that is generated from the function **generateHiddenProposal( )** by interchanging all occurrences of **hidden** and **visible** and replacing the product **visibleValuesR * weightMatrixRC** by **hiddenVariablesR * weightMatrixRC'**. This yields a new visible vector, **visibleValuesProposalR**, which we designate $\vec{v}'$. A further hidden node prediction, $\vec{h}'$ termed **hiddenValuesProposalR** in the program is then generated from $\vec{v}'$. Since a priori $\vec{h}$ constitutes the optimal set of hidden variables, any elements or products of pairs of elements of $\vec{v}'$ and $\vec{h}'$ that do not coincide with those of $\vec{v}$ and $\vec{h}$ can be considered as error terms. Accordingly, to amplify the terms coming from $\vec{v}$ and $\vec{h}$ while simultaneously diminishing those arising from $\vec{v}'$ and $\vec{h}'$, a small term proportional to $\vec{v} \otimes \vec{h} - \vec{v}' \otimes \vec{h}'$, namely **positiveWeightGradient – negativeWeightGradient** in the program, is added to $(\mathbf{M})_{ij}$. The constant of proportionality, $\lambda$, is termed the **learningRate**. Similarly, the weights and biases are corrected by adding the terms $\lambda(\vec{h} - \vec{h}')$ and $\lambda(\vec{v} - \vec{v}')$ to $\vec{b}_h$ and $\vec{b}_v$, respectively. Note that the since the inputs are either 0 or 1, the only magnitudes in the model are the initial biases and weights, governed by **upperBiasValue** and **lowerBiasValue** (as well as possibly **inverseTemperature**). A very small ratio of **learningRate** to the remaining quantities therefore results in the model only incorporating the average of the training data and therefore neglecting fluctuations among the individual samples, as will be demonstrated in the subsequent section.

An alternative formalism differs from the above procedure in that the vectors $\vec{h}_r$, $\vec{v}_r$ and $\vec{h}'_r$ are employed to calculate the updated weights and biases; that is, for example, $\vec{v} \otimes \vec{h} - \vec{v}' \otimes \vec{h}'$ is replaced by $\vec{v} \otimes \vec{h}_r - \vec{v}' \otimes \vec{h}'_v$. [33] This removes statistical fluctuations in the gradient that arise from the stochastic quantization of the nodal excitations. The body of the **for** loop above is therefore replaced by:

```
computedHiddenProbabilitiesR = fermiFunction( visibleValuesR * weightMatrixRC + …
    hiddenBiasesR, inverseTemperature );

hiddenValuesProposalR = zeros( 1, numberOfHiddenNodes );
randomHiddenValuesR = rand( 1, numberOfHiddenNodes );
hiddenValuesProposalR(computedHiddenProbabilitiesR > randomHiddenValuesR) = 1;

computedActiveProbabilitiesR = fermiFunction( hiddenValuesProposalR * weightMatrixRC' + …
    visibleBiasesR, inverseTemperature );

visibleValuesProposalR = zeros( 1, numberOfVisibleNodes );
randomVisibleValuesR = rand( 1, numberOfVisibleNodes );
visibleValuesProposalR(computedActiveProbabilitiesR > randomVisibleValuesR) = 1;

computedHiddenProbabilitiesNewR = fermiFunction( visibleValuesProposalR * ...
    weightMatrixRC + hiddenBiasesR, inverseTemperature );

positiveWeightGradientRC = visibleValuesR' * computedHiddenProbabilitiesR;
negativeWeightGradientRC = visibleValuesProposalR' * computedHiddenProbabilitiesNewR;

differenceOfWeightsRC = positiveWeightGradientRC - negativeWeightGradientRC;
differenceOfVisibleValuesR = visibleValuesR - visibleValuesProposalR;
differenceOfHiddenValuesR = computedHiddenProbabilitiesR - ...
    computedHiddenProbabilitiesNewR;

weightMatrixRC = weightMatrixRC + learningRate * differenceOfWeightsRC;
visibleBiasesR = visibleBiasesR + learningRate * differenceOfVisibleValuesR;
hiddenBiasesR = hiddenBiasesR + learningRate * differenceOfHiddenValuesR;
```

Once the restricted Boltzmann machine is trained, its accuracy can be established, as observed in [28] [29] by employing it to regenerate known properties of the original training sequence. The code for this "block Gibbs sampling" step is very compact (but observe that if memory space is not preallocated in MATLAB for the output data as below the performance is severely degraded). The synthetic data samples output by the procedure are stored in the **numberOfRealizations** rows of the matrix **computedVisiblesR**.

```
visibleValuesR = randi( 2, 1, numberOfVisibleNodes ) - 1;
computedVisiblesR = zeros( numberOfRealizations, numberOfVisibleNodes );

for loop = 1 : numberOfRealizations
   for innerLoop = 1 : numberOfBlockIterations
      hiddenProposalR = generateHiddenProposal( visibleValuesR, weightMatrixRC, ...
         hiddenBiasesR, numberOfHiddenNodes, inverseTemperature );
      visibleValuesR = generateVisibleProposal( hiddenProposalR, weightMatrixRC, ...
         visibleBiasesR, numberOfVisibleNodes, inverseTemperature );
   end
   computedVisiblesR(loop, :) = visibleValuesR(:);
end
```

This strategy assumes that the statistical properties of the restricted Boltzmann machine are reflective of those of the training data. The sample fluctuations are imparted by the random number generators within **generateVisibleProposal( )** and **generateHiddenProposal( )**.

**Results:** To quantify the accuracy of the RBM technique requires a detailed analysis of the influence of its hyperparameters on the statistics of the block sampling output. As a first step, a problem must be identified that is highly sensitive to algorithmic features. In earlier studies of transition matrix based biased sampling procedures, specific heat computations at temperatures for which a large number of states are spread evenly throughout a wide region of energy-magnetization space satisfied this criterion. [12] [14] In this paper, we accordingly consider the two-dimensional $8 \times 8$ spin Ising model example analyzed in [28] [29] [30] at such a temperature. In particular, the input data consists of 600,000 samples generated with the Metropolis Monte-Carlo algorithm for a temperature of $T/k_b = 3.526$. The input samples are characterized by the probability distribution of Figure 1, where the magnetization and the energy, in normalized temperature and energy units, $T \equiv T/k_b$ and $E \equiv E/J$, where $J$ is the spin-spin coupling constant, are displayed on the horizontal and vertical axes, respectively. The specific heat per unit spin, calculated from the energy fluctuations among the different samples according to

$$c = -\frac{<E>^2 - <E^2>}{T^2}$$

equals 0.2599 for the samples employed, which compares to the exact result, $c = 0.2556$. [34] This input sequence is employed to train a RBM with 64 active and 64 hidden nodes. Since the results depend on the random number sequence, two sets of studies were undertaken. In the first of these, the random number sequence was fixed so that the influence of various changes to the RBM could be investigated. Subsequently, a single set of hyperparameter values were instead employed but the variations associated with different sets of random numbers and number of hidden nodes were examined. The calculations were, unless otherwise specified, performed with the first of the two implementations of the RBM discussed above as it is more commonly encountered in the broader machine learning literature. The second procedure generates similar if somewhat more accurate results, as demonstrated below.

Accordingly, a single random number sequence was first specified through the statement **rng( 1 )**. For this sequence, the output of a single run of the program, which requires 5 minutes on a Ryzen 7 2700 processor and returns $c = 0.2775$, is shown in Figure 2 for **upperBiasValue = --lowerBiasValue = 0.02, inverseTemperature = 1.0** and **learningRate = 0.001**. The thin dashed line in this and succeeding figures reproduces the result of Figure 1 for comparison. Observe that for this random number sequence the maximum of the Boltzmann machine distribution is displaced slightly toward higher magnetizations while its tail is somewhat elongated toward lower magnetizations. The asymmetry can however be eliminated by setting **inverseTemperature** to value larger than unity; for example when **inverseTemperature = 1.5**, the state distribution becomes effectively symmetric, c.f. Figure 3, and the specific heat value falls to 0.2299. Alternatively the training procedure itself can be modified so that, for example the constrastive divergence step becomes **differenceRC = differenceRC + positiveGradRC – negativeFactor * negativeGradRC** with a similar modification of the constrastive divergence expression for the bias terms**.** Setting **negativeFactor = 1.007**, for example, leads to the distribution of Figure 4 with $c = 0.2798$.

A further impact on the distribution arises from the choice of **learningRate**. A learning rate of 0.0001 yields a more compact, stable and circular state distribution than the physical initial distribution, c.f. Figure 5, with a greatly improved specific heat value of 0.2499. It should be noted however that the statistical nature of the calculation implies that for different random number sequences the specific heat fluctuates among consecutive results with accompanying changes in the energy-magnetization distribution. For example, Figure 6 shows the result for the specific heat as a function of the learning rate from $1.0 \times 10^{-5}$ to $1.0 \times 10^{-3}$ in steps of $1.0 \times 10^{-5}$. While the average of the curve is effectively constant except at the smallest values of learning rate shown, considerable fluctuations exist around the correct value that increase with learning rate. This implies a tradeoff between accuracy in the determination of the specific heat (which requires a small learning rate) and the fidelity with which the correct distribution of the realizations in energy-magnetization space is reproduced (as the distribution is better modeled for large learning rates). The calculation of Figure 6 is repeated in Figure 7 but with **upperBiasValue = – lowerBiasValue= 0.25**, demonstrating that the qualitative behavior of the algorithm is relatively unaffected by the initial conditions. However, the specific heat is somewhat improved in accuracy so that these upper and lower bias values are employed in subsequent calculations. The corresponding result for the second procedure of the preceding section is given in Figure 8. This curve displays similar features but improved statistics compared to the first procedure as is expected from the elimination of the quantization step in the evaluation of the biases and weights.

A similar situation exists for the specific heat as a function of the inverse temperature of for a learning rate $1.0 \times 10^{-3}$ and **upperBiasValue= –lowerBiasValue = 0.25**, c.f. Figure 9, which displays the variation of the specific heat with **inverseTemperature** for values spaced by 0.01 between 1.0 and 1.5. The specific heat again fluctuates markedly but is on average insensitive to this parameter. Again, the fluctuations again can be significantly suppressed by decreasing the learning rate to $1.0 \times 10^{-4}$ and employing the second form of the RBM, as illustrated in Figure 10.

The fluctuations in the specific heat as a function of the various control parameters when the random number sequence is fixed are reflected in the results for the specific heat calculations for different random number sequences; that is, when the statement **rng( 1 )** is absent. To demonstrate, we repeat the calculation of Figure 2 one hundred times and display the specific heat values as a histogram in Figure 11. Again, the distribution in energy-magnetization space varies widely among realizations with the degree of distortion typically closely related to the error in $c$. For a smaller value of **learningRate = 0.0001**, the fluctuations, as expected, are greatly reduced, as evidenced in Figure 12, but the energy-magnetization distribution then as before departs significantly from that of Figure 1. The analogous results for the second method of the previous section, Figure 13 and Figure 14, while similar in form to those generated by the first method, as expected display significantly smaller statistical errors. However, the associated probability distribution functions on average differ only marginally from Figure 2 and Figure 5.

Finally, if the number of hidden nodes is decreased to **numberOfHiddenNodes = 8**, the histogram corresponding to Figure 11 is given in Figure 15 for a learning rate of $1.0 \times 10^{-4}$ and in Figure 16 for a rate of $1.0 \times 10^{-3}$. Evidently, reducing the number of hidden nodes and therefore the number of free parameters in the RBM considerably decreases the histogram width since far fewer samples are required before the model attains an optimal or near-optimal state. Indeed, the energy-magnetization distribution, of Figure 17 for the $1.0 \times 10^{-3}$ learning rate corresponding to Figure 2, while displaced upward in energy, displays a more compact and symmetric structure. This is associated with a greater degree of optimization after training and consequently less variation during block Gibbs sampling. As well, the specific heat value associated with this calculation, 0.2655, also displays an enhanced accuracy compared to the corresponding calculation with 64 hidden nodes, which presumably again follows from far smaller, if unphysical, spread of the specific heat histogram. However, the sharp decline of the specific heat at the smallest learning rates evident in Figure 6 - Figure 8 is displaced to higher values by approximately an order of magnitude for 8 hidden nodes, as is clearly evident from the markedly lower average specific heat in Figure 17.

**Conclusions:** This paper has examined the precision of the RBM applied to the two-dimensional Ising model. Deviations from the properties of the training set and from exact results are evident not only in the specific heat, which fluctuates as the model parameters and random number sequence are varied, but also in the joint energy and magnetization distribution of the generated samples. While the statistical fluctuations in the model predictions are highly dependent on the number of hidden nodes and the values of the various hyperparmeters, a clear tradeoff exists between the magnitude of the fluctuations in the specific heat and the fidelity with which the RBM reproduces the $E-H$ distribution of the input realizations. The optimal values of model parameters such as the learning rate and the number of hidden nodes therefore depend inherently on the nature of the statistical quantity or quantities of interest.

**Acknowledgements:** Ejaaz Merali is particularly acknowledged for important discussions and for numerical confirmation of the results in this work. This work was supported by the Natural Sciences and Engineering Research Council of Canada (NSERC) is acknowledged for financial support. Additionally one of us (RGM) is further funded by the Canada Research Chair (CRC) program, and the Perimeter Institute for Theoretical Physics. Research at Perimeter Institute is funded in part by the Government of Canada through the Department of Innovation, Science and Economic Development Canada and by the Province of Ontario through the Ministry of Colleges and Universities.

**Figures:**

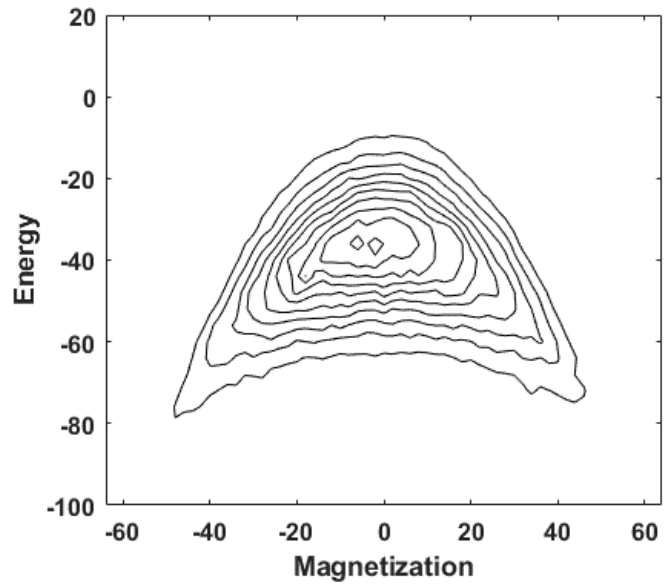

*Figure 1: The joint energy-magnetization distribution of the 60,000 input Metropolis samples for a $8 \times 8$ two-dimensional Ising model*

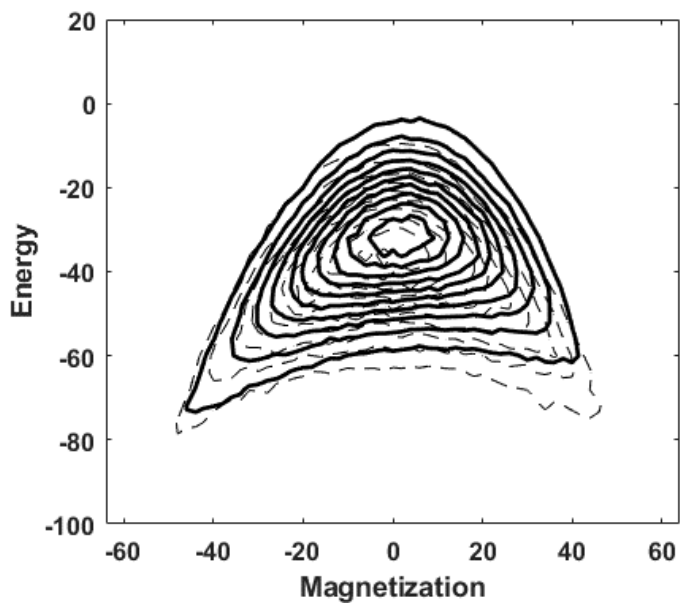

*Figure 2: A comparison of the energy magnetization distribution generated by the RBM for 64 hidden nodes in a calculation with bias values = 0.02, inverse temperature = 1 and learning rate $1.0 \times 10^{-3}$ (solid line) with the input distribution of Figure 1 (dashed line)*

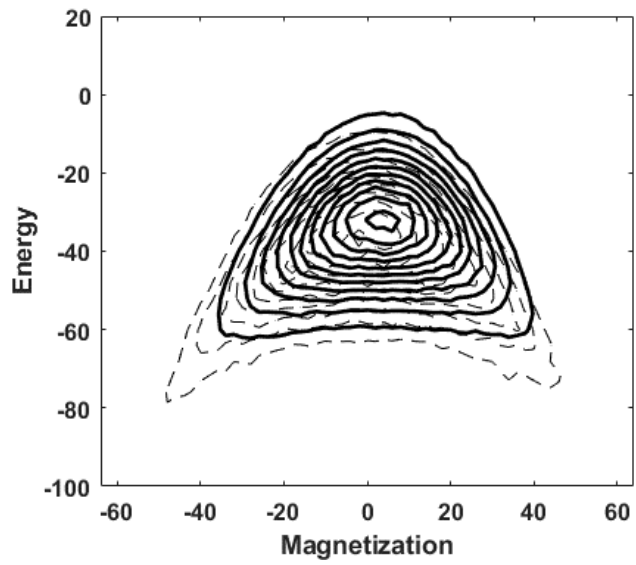

*Figure 3: As in Figure 2, but with **inverseTemperature = 1.5***

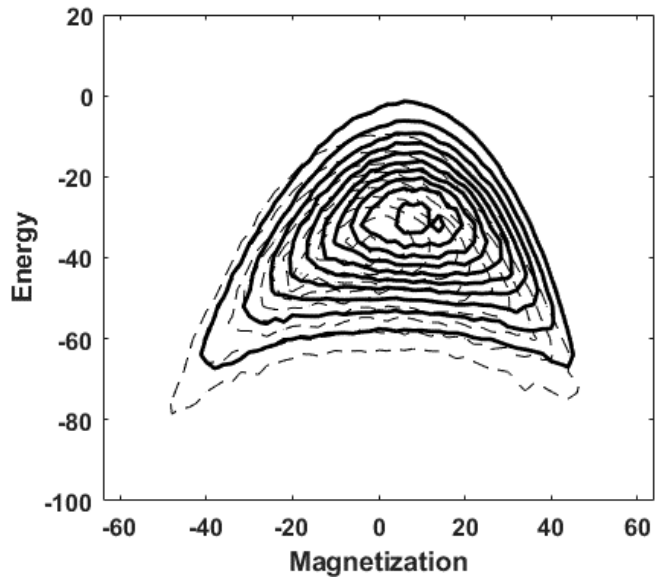

*Figure 4: As in Figure 2, but with a modified contrastive divergence step in the learning phase of the RBM.*

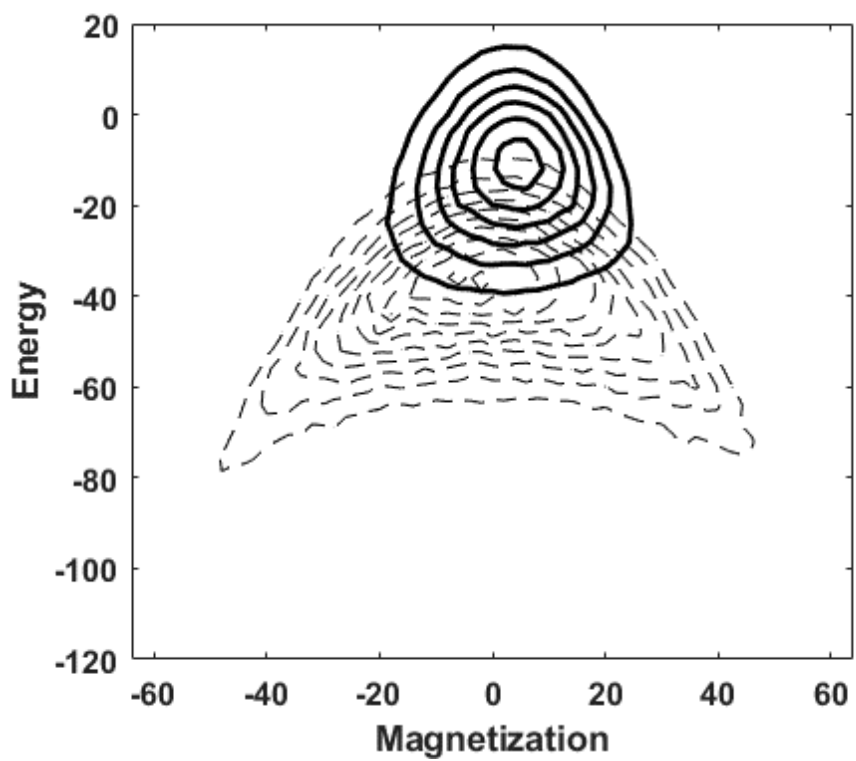

Figure 5: As in Figure 2 but with a learning rate of $1.0 \times 10^{-4}$.

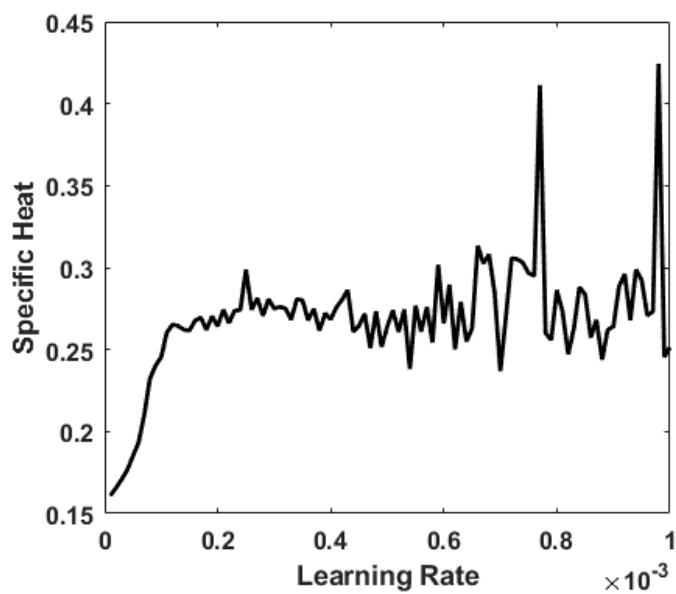

Figure 6: The specific heat for the calculation of Figure 2 as the learning rate is varied from $1.0 \times 10^{-5}$ to $1.0 \times 10^{-3}$ in steps of $1.0 \times 10^{-5}$.

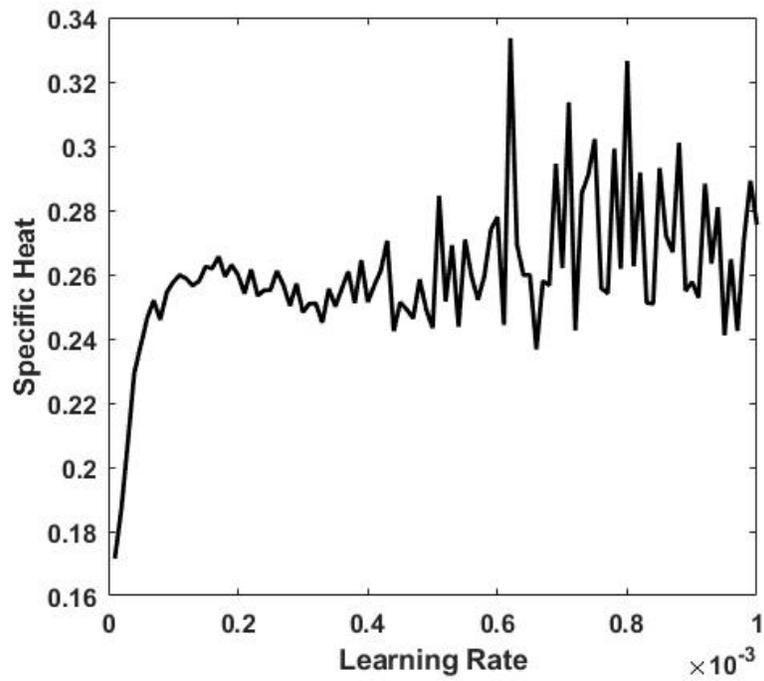

*Figure 7: As in Figure 6 but with initial bias values of $\pm 0.25$*

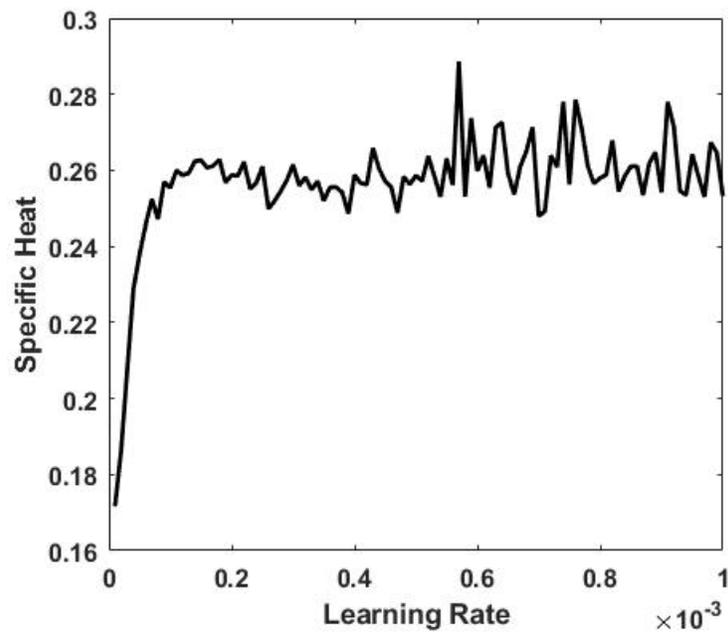

*Figure 8: As in Figure 6 but for the modified RBM technique with initial bias limits of $\pm 0.25$*

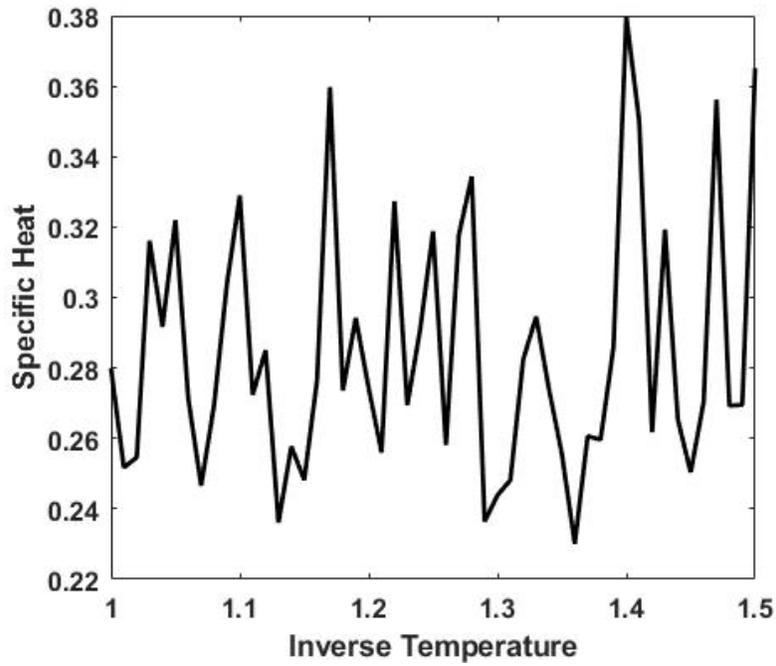

Figure 9: The variation of the specific heat with the inverse temperature for the calculation of Figure 2 and initial bias limits between f $\pm 0.25$.

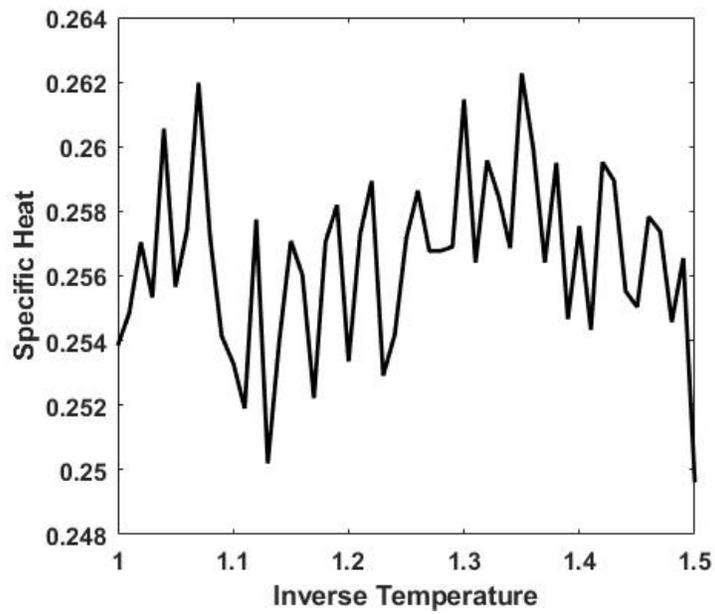

Figure 10: The specific heat as a function of the inverse temperature parameter for initial bias limits of $\pm 0.25$ and a learning rate of $1 \times 10^{-4}$ for the alternative (second) RBM technique.

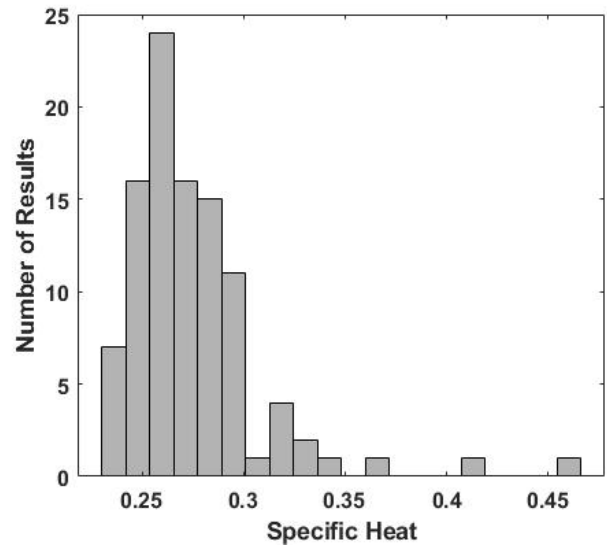

*Figure 11: A histogram of 100 results for the specific heat for the calculation of Figure 2 with different random number sequences*

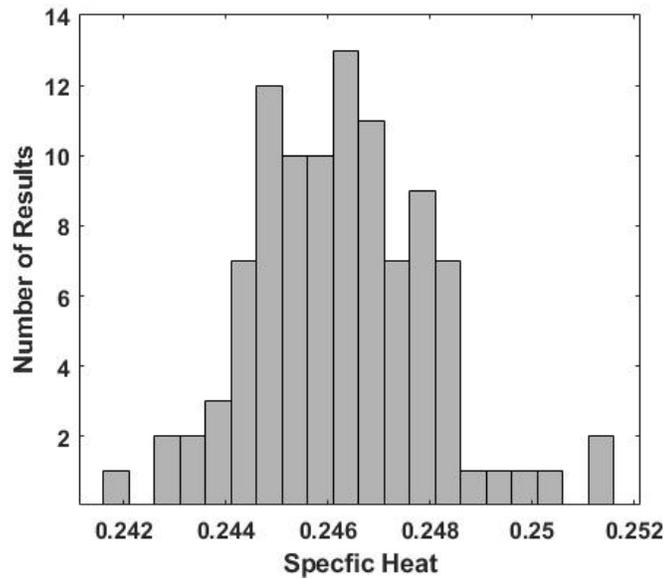

*Figure 12: As in the previous figure, except for a learning rate of $1.0 \times 10^{-4}$.*

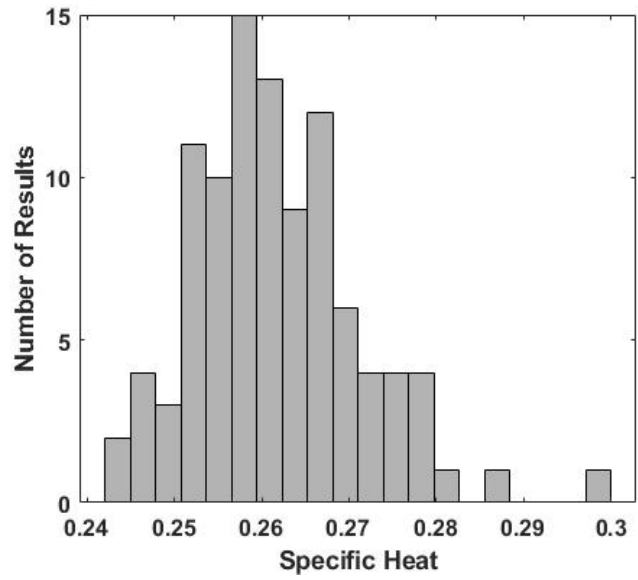

*Figure 13: As in Figure 11, but for the alternative RBM formalism in the text*

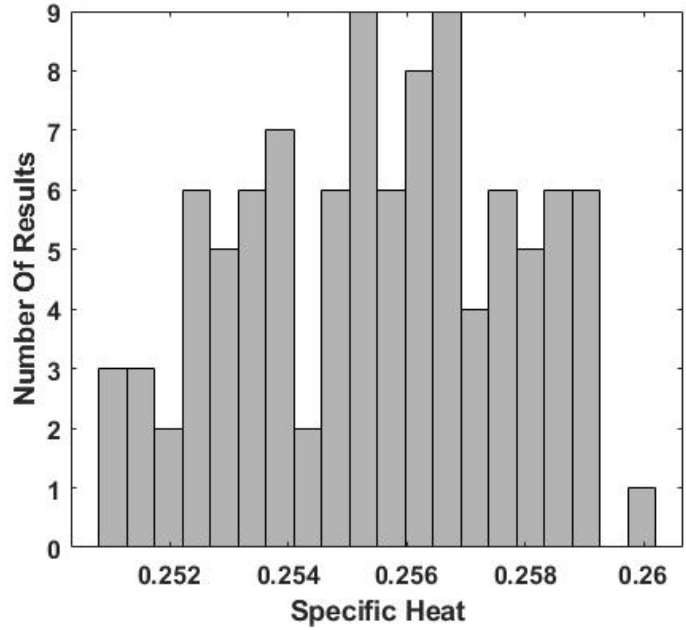

*Figure 14: As in Fig. 11, but for the alternative RBM formalism in the text*

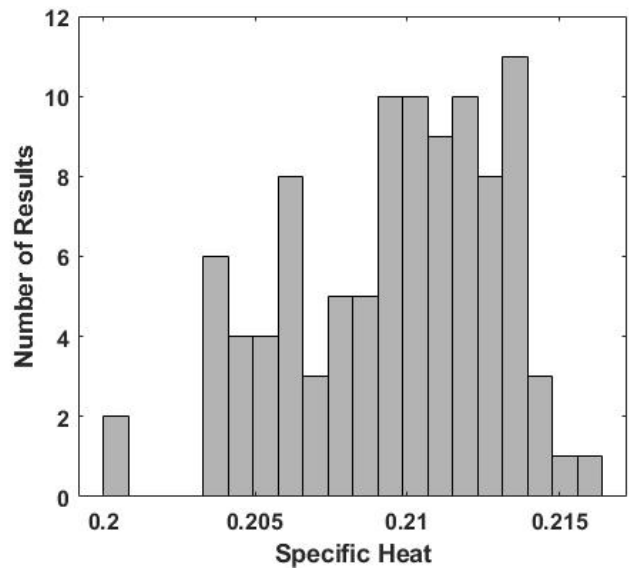

*Figure 15: As in Fig. 11, but for 8 hidden nodes and a learning rate of $1.0 \times 10^{-4}$.*

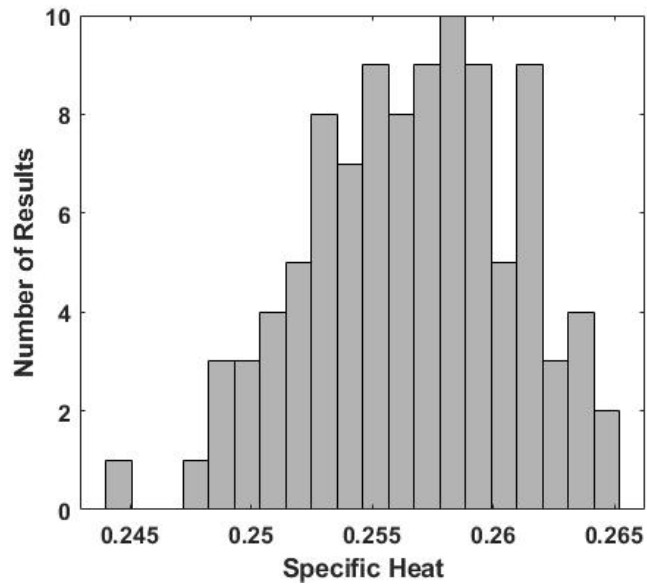

*Figure 16: As in Fig. 11, but for 8 hidden nodes and a learning rate of $1.0 \times 10^{-3}$.*

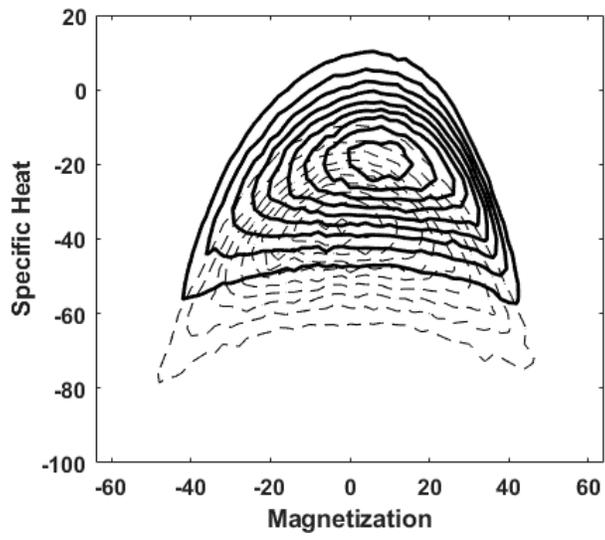

*Figure 17: The energy-magnetization diagram of Figure 2, but with 8 rather than 64 hidden nodes.*